\documentclass[a4paper,10pt]{iopart}
\usepackage{graphics}
\usepackage{graphicx}
\usepackage{epsfig}

\usepackage{amssymb}

\newcommand{\Teff}{T_{\mbox{\small{\textit{eff}}}}}
\newcommand{\pT}{p_{\mbox{\small{\textit{T}}}}}
\begin{document}
%opening
\title{NA60 results on $p_T$ spectra and the $\rho$ spectral function in In-In collisions}
\author{J. Seixas$^{5}$ for the NA60 Collaboration:\footnote{This work was supported in part by FCT under grant POCI/FP/63919/2005}
}
\vspace*{-10pt}
\author{R.~Arnaldi$^{10}$, 
R.~Averbeck$^{9}$, 
K.~Banicz$^{2,4}$, 
J.~Castor$^{3}$,
%\author{
B.~Chaurand$^{7}$,
C.~Cical\`o$^{1}$
A.~Colla$^{10}$, 
P.~Cortese$^{10}$,
%\author{
S.~Damjanovic$^{2}$,
A.~David$^{2,5}$,
A.~De~Falco$^{1}$, 
A.~Devaux$^{3}$,
%\author{
A. Drees$^{9}$,
L.~Ducroux$^{6}$,
H.~En'yo$^{8}$, 
A.~Ferretti$^{10}$, 
M.~Floris$^{1}$,
%\author{
A.~F\"{o}rster$^{2}$,
P.~Force$^{3}$,
N.~Guettet$^{3}$,
A.~Guichard$^{6}$, 
H.~Gulkanian$^{11}$,
%\author{J.~Heuser$^{8}$,
M.~Keil$^{2,5}$,
L.~Kluberg$^{7}$, 
C.~Louren\c{c}o$^{2}$, 
J.~Lozano$^{5}$,
%\author{
F.~Manso$^{3}$, 
A.~Masoni$^{1}$,
P.~Martins$^{2,5}$, 
A.~Neves$^{5}$, 
H.~Ohnishi$^{8}$, 
%\author{
C.~Oppedisano$^{10}$,
P.~Parracho$^{2}$, 
P.~Pillot$^{6}$, 
G.~Puddu$^{1}$,
%\author{
E.~Radermacher$^{2}$,
P.~Ramalhete$^{2}$, 
P.~Rosinsky$^{2}$, 
E.~Scomparin$^{10}$,
%\author{
%J.~Seixas$^{5}$,
S.~Serci$^{1}$, 
R.~Shahoyan$^{2,5}$, 
P.~Sonderegger$^{5}$,
%\author{
H.J.~Specht$^{4}$,
R.~Tieulent$^{6}$, 
G.~Usai$^{1}$, 
R.~Veenhof$^{5,2}$, 
H.K.~W\"ohri$^{1}$}

\address{
$^{1}$Univ.\ di Cagliari and INFN, Cagliari, Italy,
$^{2}$CERN, Geneva, Switzerland,
$^{3}$LPC, Univ.\ Blaise Pascal and CNRS-IN2P3, Clermont-Ferrand, France,
$^{4}$Univ.\ Heidelberg, Heidelberg, Germany,
$^{5}$LIP, Lisbon, Portugal,
$^{6}$IPN-Lyon, Univ.\ Claude Bernard Lyon-I and CNRS-IN2P3, Lyon, France,
$^{7}$LLR, Ecole Polytechnique and CNRS-IN2P3, Palaiseau, France,
$^{8}$RIKEN, Wako, Saitama, Japan,
$^{9}$SUNY Stony Brook, New York, USA,
$^{10}$Univ.\ di Torino and INFN, Italy,
$^{11}$YerPhI, Yerevan, Armenia
}

%\maketitle

\begin{abstract}
The NA60 experiment at the CERN SPS has studied low-mass muon pairs in 158 AGeV In-In
collisions. A strong excess of pairs is observed above the yield expected from neutral meson decays.
The unprecedented sample size of close to 400K events and the good mass resolution of about 2\%
have made it possible to isolate the excess by subtraction of the decay sources (keeping the $\rho$). The
shape of the resulting mass spectrum exhibits considerable broadening, but essentially no shift in mass. The
acceptance-corrected transverse-momentum spectra have a shape atypical for radial flow and
show a significant mass dependence, pointing to different sources in different mass regions. 
\end{abstract}
\vspace*{-20pt}
\section{Introduction}
Thermal dilepton production in the mass region below 1 GeV/c$^{2}$ is
largely mediated by the light vector mesons $\rho$, $\omega$ and
$\phi$. Among these, the $\rho$(770 MeV/c$^{2}$) is the most
important, due to its strong coupling to the $\pi\pi$ channel and its
short lifetime of only 1.3 fm/c. Changes both in width and in mass were
originally suggested as precursor signatures of the chiral
transition~\cite{Pisarski:mq} and subsequent models have tied these changes
directly~\cite{Brown:kk} or
indirectly~\cite{Rapp:1999ej} to chiral symmetry restoration.

The first NA60 results focused on the space-time averaged
spectral function of the $\rho$ \cite{Arnaldi:2006}; more details were
added recently~\cite{hq:2006}. The present paper
concentrates mainly on new developments, in particular first results on
acceptance-corrected $p_{T}$ and $m_{T}$ spectra for different mass
windows
\section{Experimental results}
\label{sec:1}
\begin{figure*}[t!]
\centering
\includegraphics*[width=5.28 cm, height=4.65 cm ]{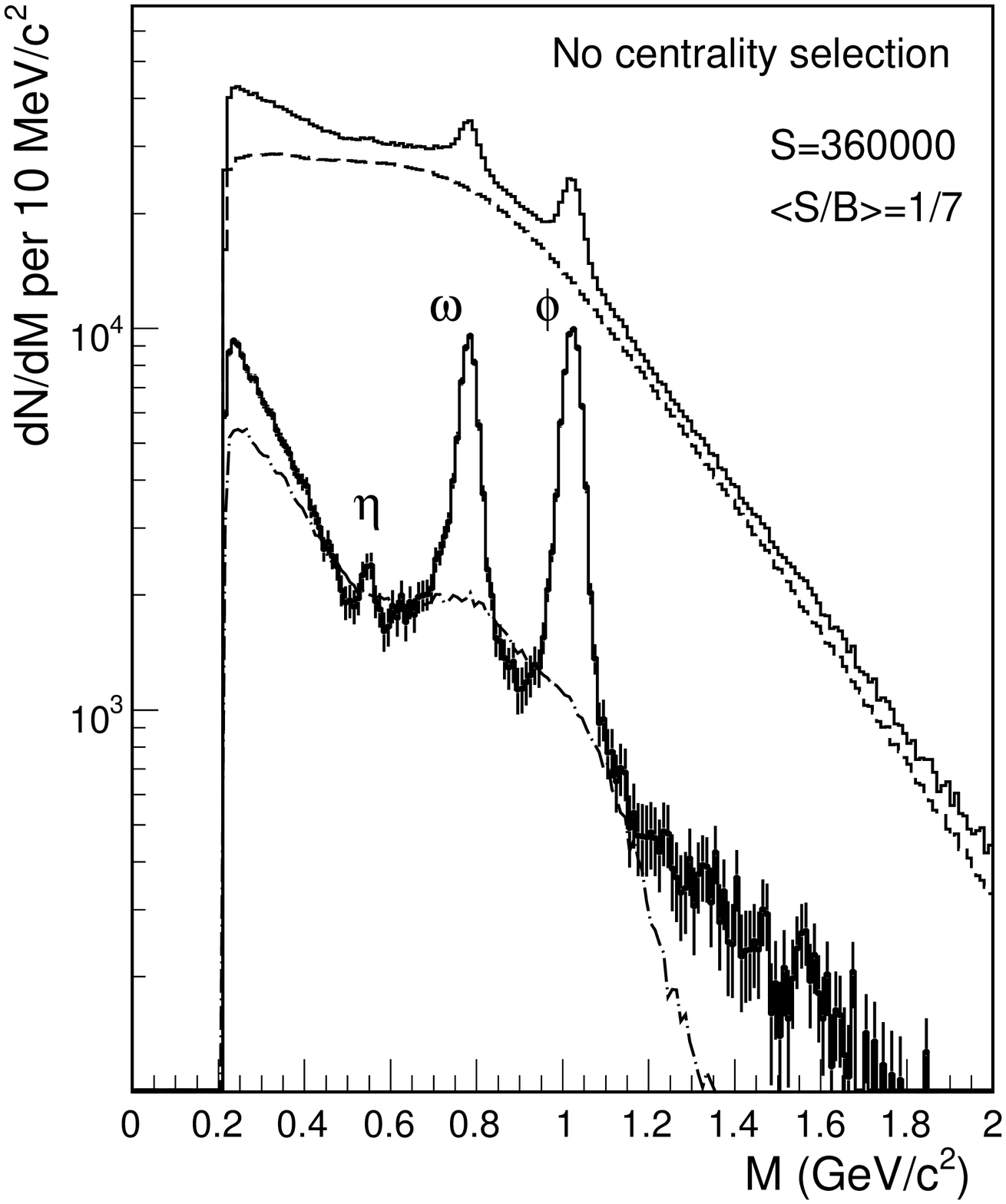}%,clip=, bb = 6 12 560 665]{img-final/fig1_a.eps}
\includegraphics*[width=5.28 cm, height=4.65 cm ]{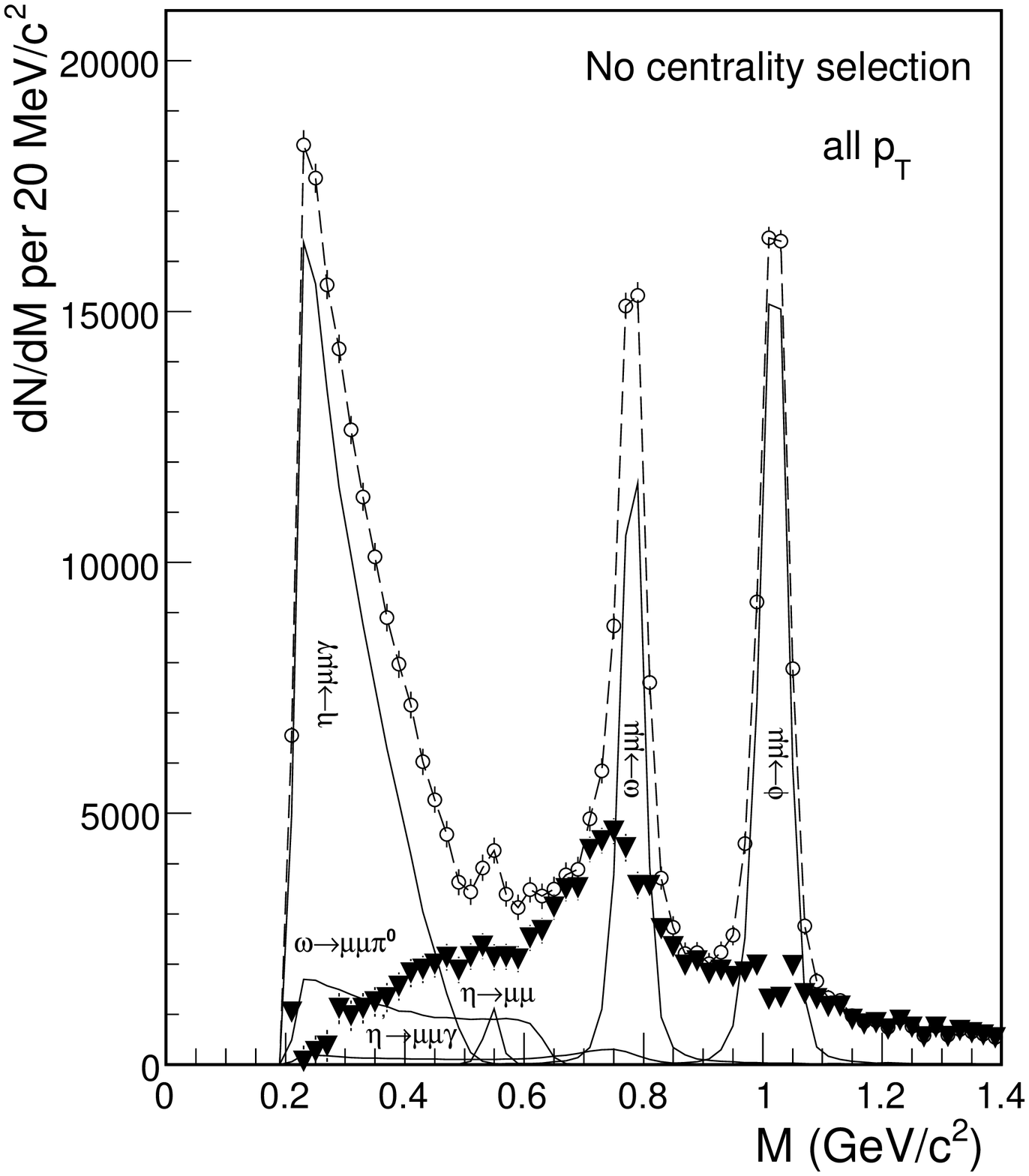}%,clip=, bb= 0 12 560 665]{img-final/fig1_b.eps}
\caption{Left: Mass spectra of the opposite-sign dimuons (upper
histogram), combinatorial background (dashed), signal fake matches
(dashed-dotted), and resulting signal (histogram with error
bars). Right:Isolation of an excess above the hadron decay cocktail. Total data (open circles), individual cocktail sources
(solid), difference data (thick triangles), sum of cocktail sources
and difference data (dashed).}
\label{fig1}
\end{figure*}

Details of the NA60 apparatus can be found
in~\cite{Gluca:2005,Keil:2005zq}, while the different analysis steps
(including the critical assessment of the combinatorial background
from $\pi$ and $K$ decays through event mixing) are described
in~\cite{Ruben:2006qm}. The results reported here were obtained from
the analysis of data taken in 2003 for In-In at 158 AGeV. The left
part of Fig.~\ref{fig1} shows the opposite-sign, background and signal
dimuon mass spectra, integrated over all collision centralities. After
subtracting the combinatorial background and the signal fake matches,
the resulting net spectrum contains about 360\,000 muon pairs in the
mass range 0-2 GeV/c$^2$, roughly 50\% of the total available
statistics. The associated average charged-particle multiplicity
density measured by the vertex tracker is $\langle dN_{ch}/d\eta\rangle$ =120.
Vector mesons $\omega$ and $\phi$ are completely resolved with a mass
resolution at the $\omega$ of 20 MeV/c$^{2}$. Most of the analysis is done in four classes of collision centrality
(defined through the charged-particle multiplicity density):
peripheral (4$-$30), semiperipheral (30$-$110), semicentral (110$-$170) and
central (170$-$240).
% figure 1
% end of figure 1

The peripheral data can essentially be described by the expected
electromagnetic decays of the neutral mesons~\cite{Arnaldi:2006,hq:2006}. This is not the case in the more
central bins, due to the presence of a strong excess. To isolate this
excess without any fits,
a novel procedure has been devised \cite{Arnaldi:2006,hq:2006,Ruben:2006qm}. 
The resulting spectrum for all
centralities and all $\pT$ shows a a peaked structure residing on a broad continuum. The same feature can be seen for each centrality bin, with a yield strongly
increasing with centrality, but remaining essentially centered around
the position of the nominal $\rho$ pole.

A more quantitative analysis of the shape of the excess mass spectra
\textit{vs.} centrality has been performed, using a finer subdivision
of the data into 12 centrality bins~\cite{hq:2006,Gluca:2005}. On the basis of this analysis it is possible to rule out that the excess shape can be accounted for by the cocktail $\rho$ residing on a broad continuum, independent of centrality. 
% figure3
\begin{figure}[h!]
\centering
\includegraphics[width=0.4\textwidth]{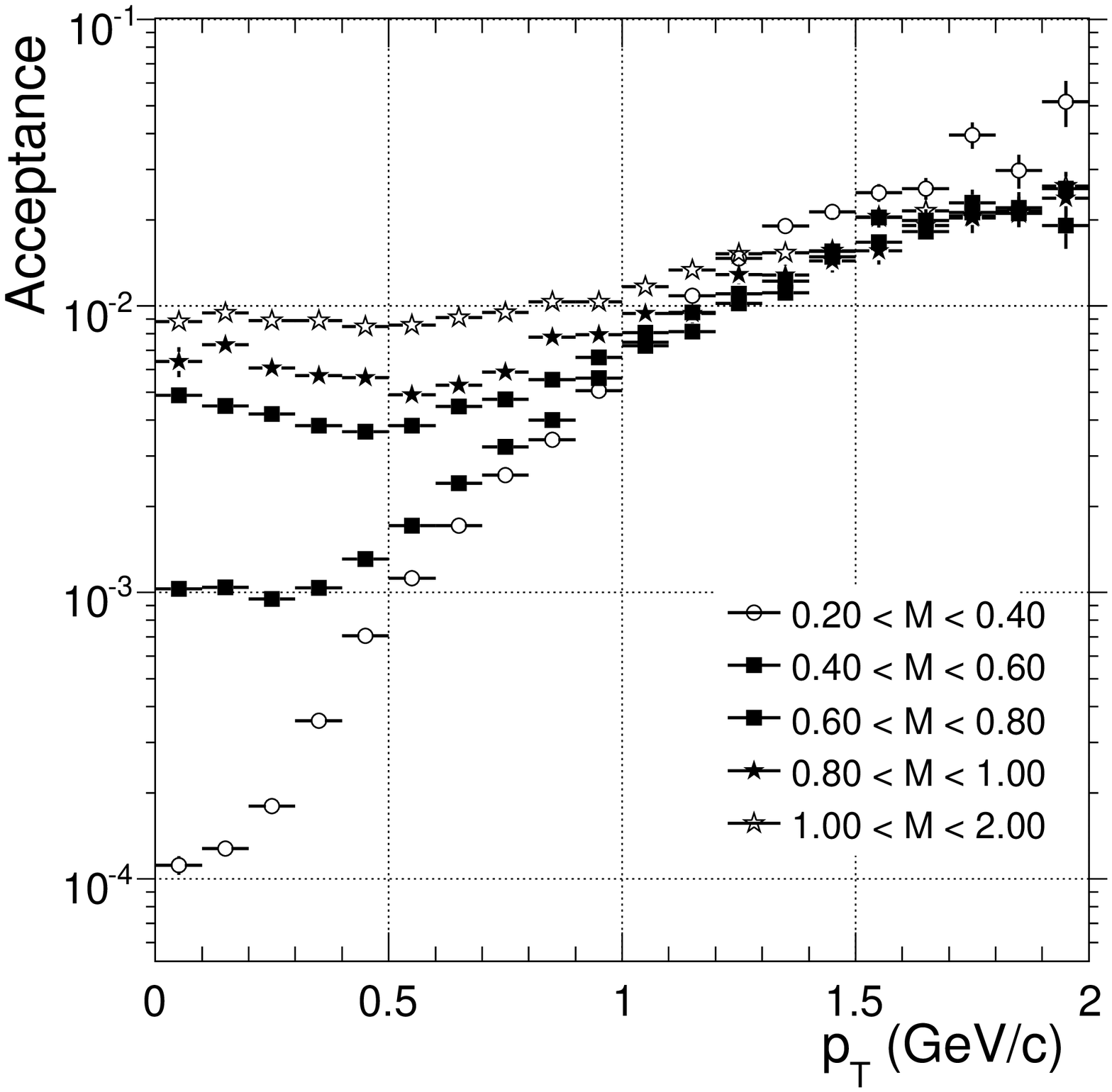} \hspace*{1cm}
\includegraphics[width=0.4\textwidth]{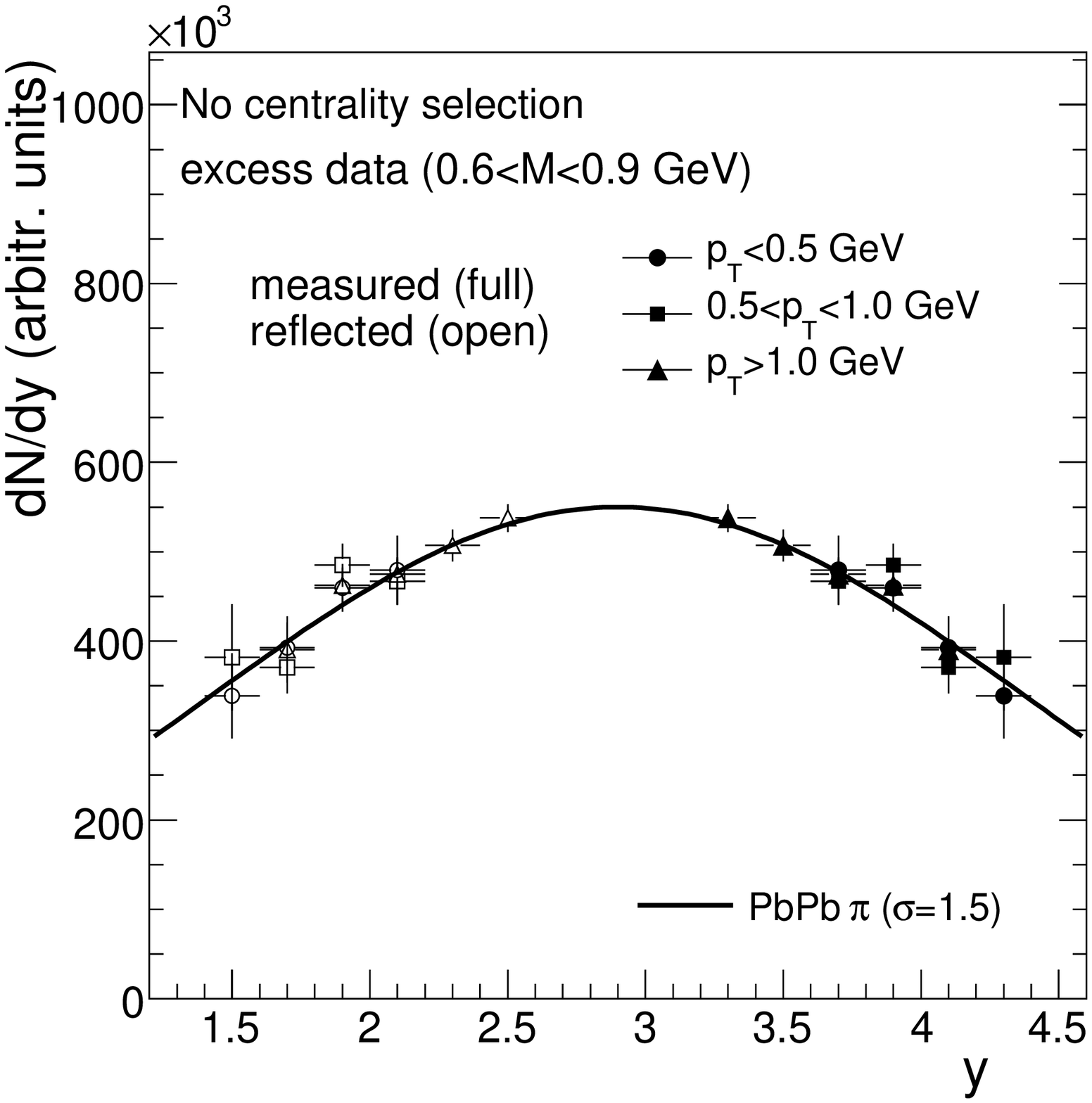}
\caption{Left: Acceptance relative to 4$\pi$ for different mass
windows. Right: Rapidity distribution of the excess data for the mass
window $0.6<\!\!M\!\!<0.9$~GeV/$c^{2}$ and for three selected $p_{T}$
bins. The measured data (full markers) are reflected around
midrapidity (open markers).}
\label{fig3}
\end{figure}
% end figure3
%figure 4
\begin{figure*}[h!]
\resizebox{\textwidth}{!}{%
\includegraphics[width=0.9\textwidth,clip]{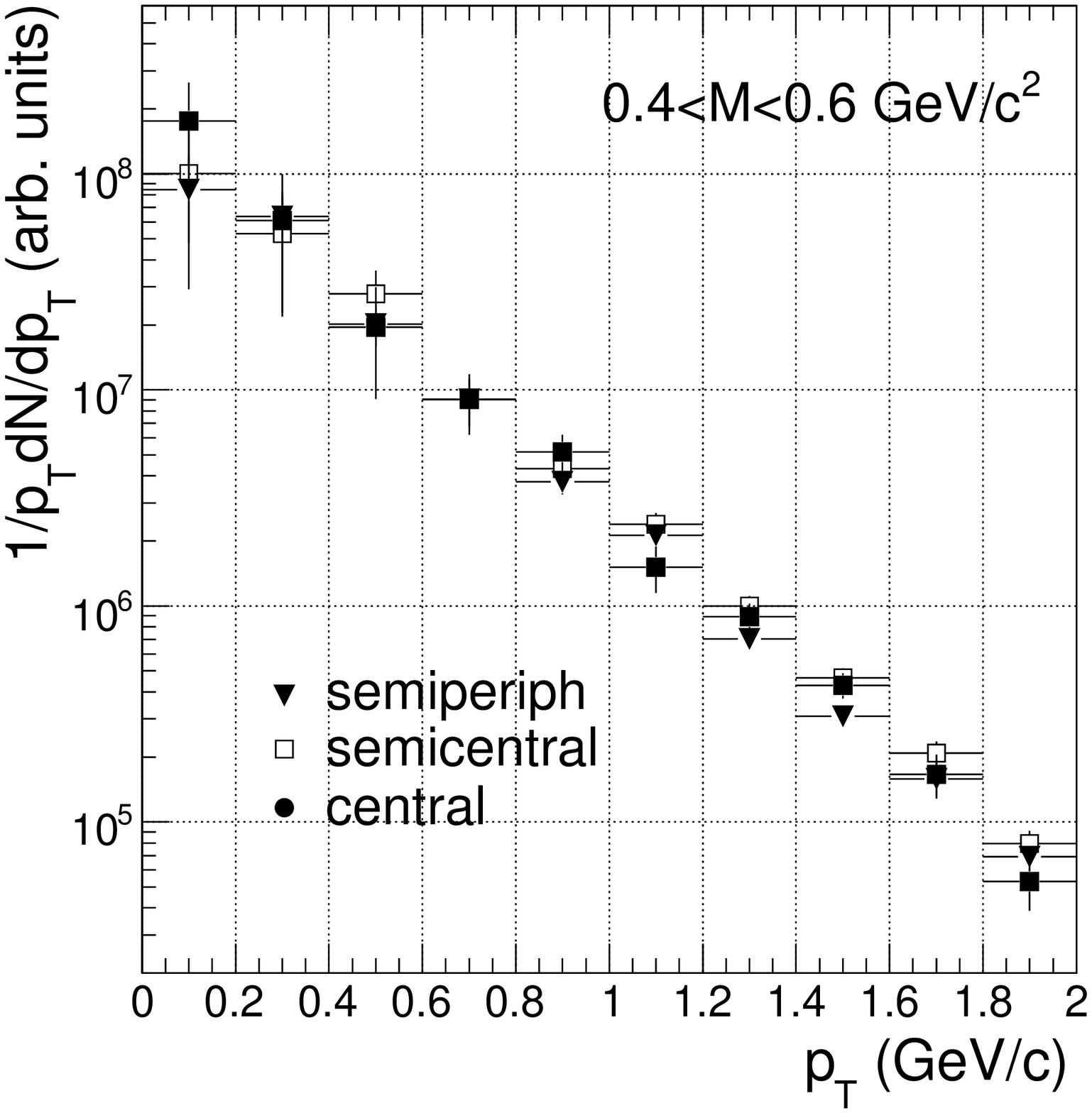}
\includegraphics[width=0.9\textwidth,clip]{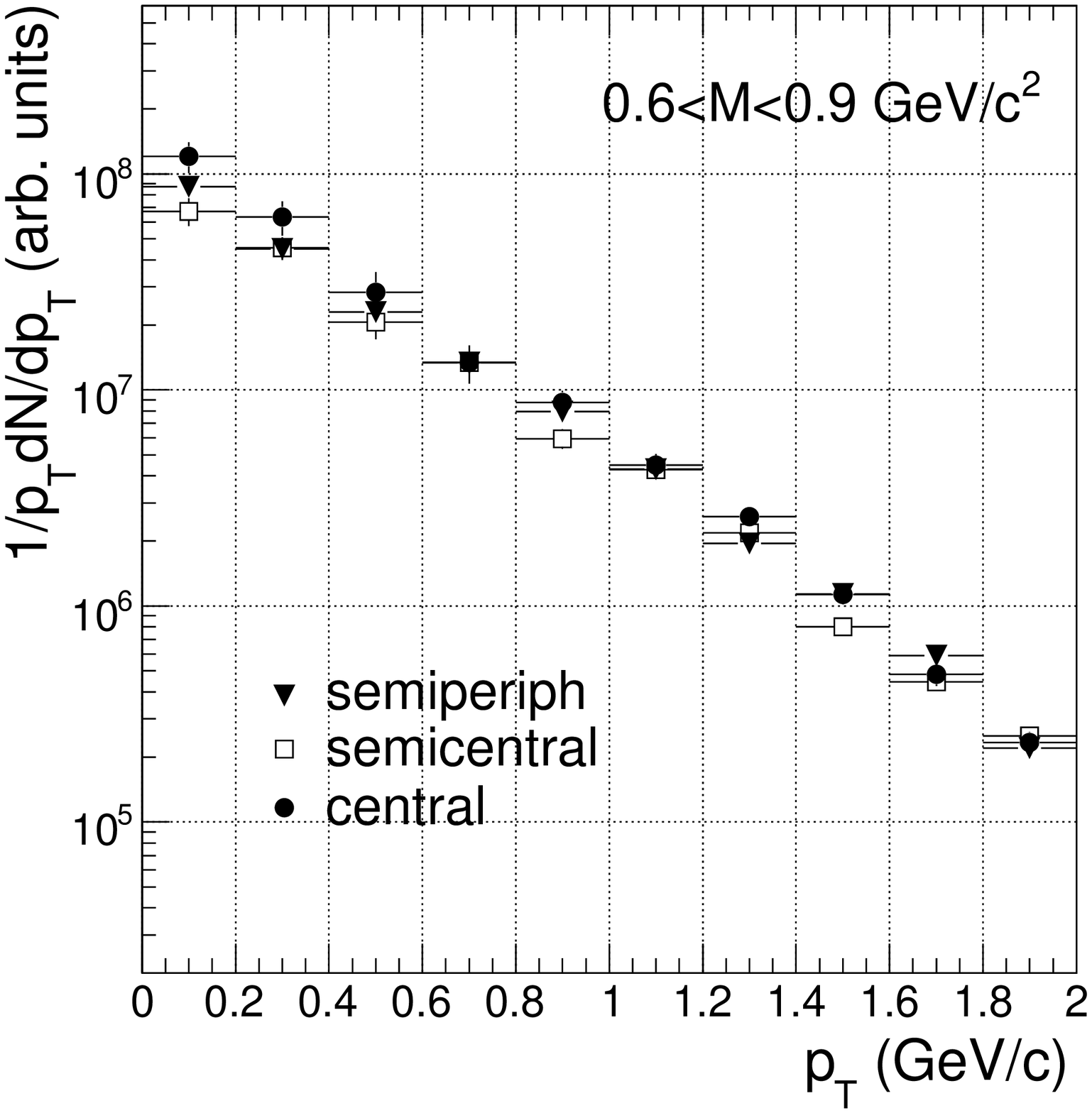}
\includegraphics[width=0.9\textwidth,clip]{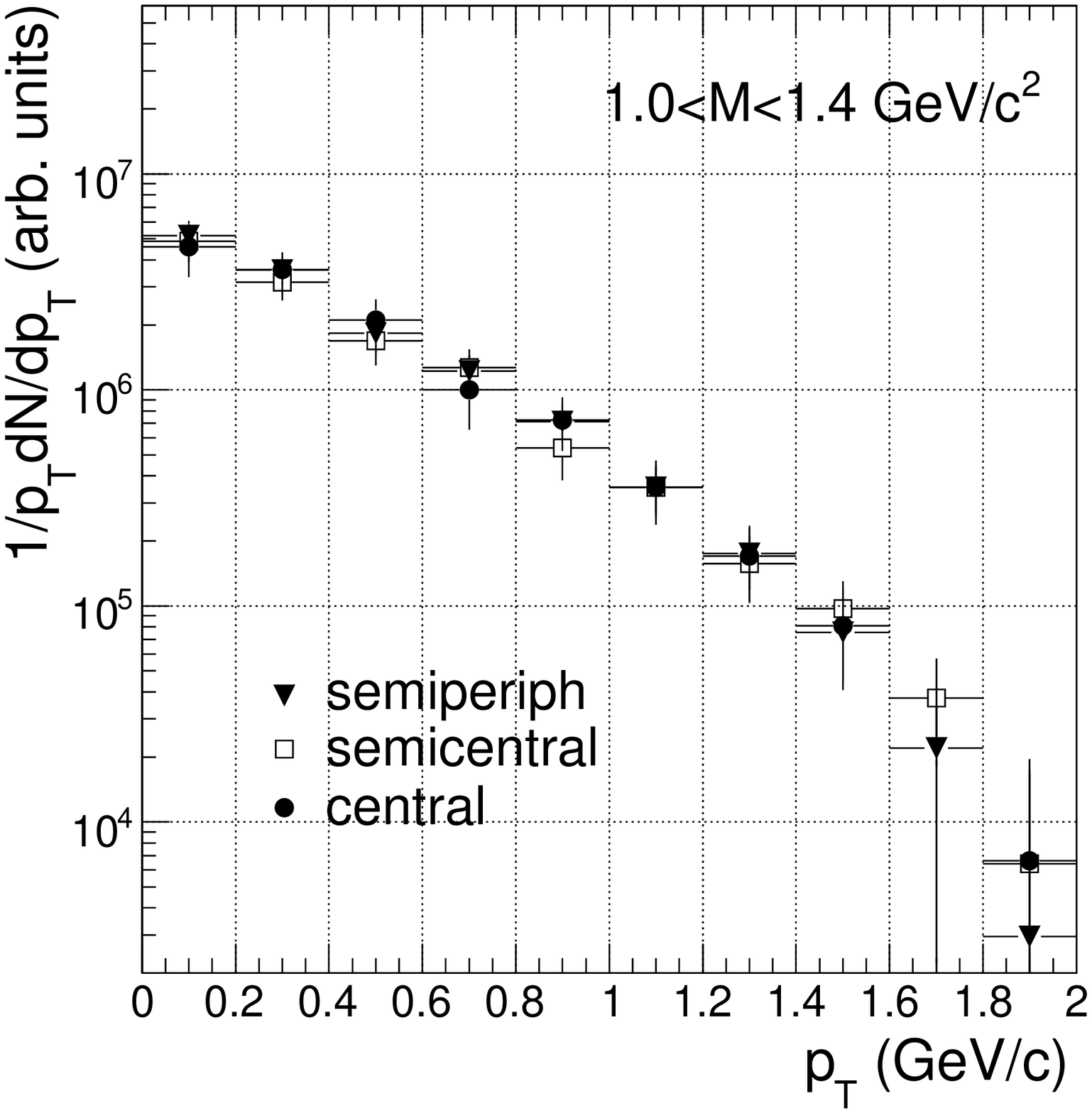}
}
\caption{Acceptance-corrected $p_{T}$ spectra for three mass windows
and for three centrality bins. For discussion of errors see text.}
\label{fig4}
\end{figure*}
%end figure 4

\begin{figure*}[h!]
\centering
\includegraphics[width=0.4\textwidth,clip]{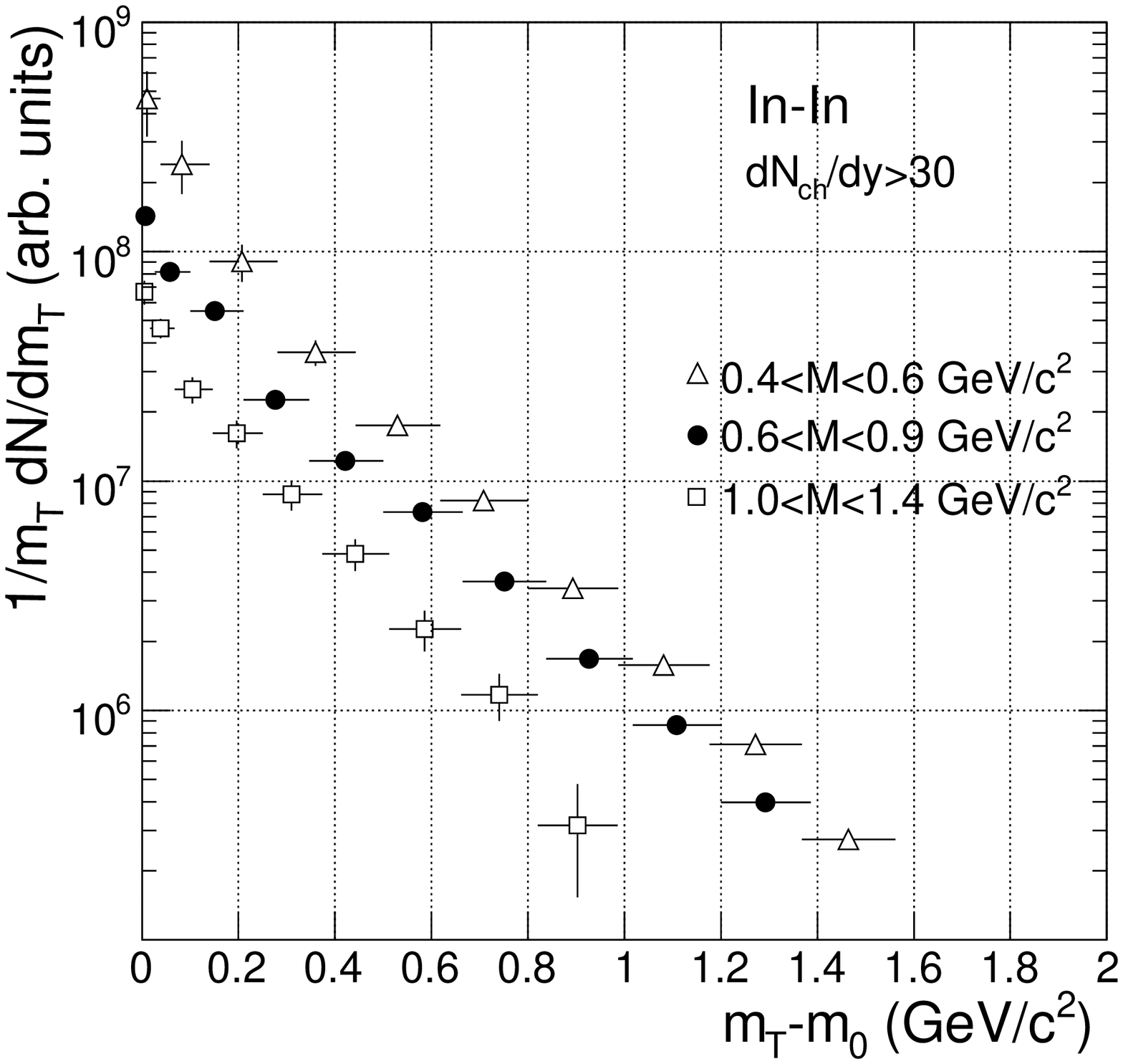}\hspace*{1cm}
\includegraphics[width=0.4\textwidth,clip]{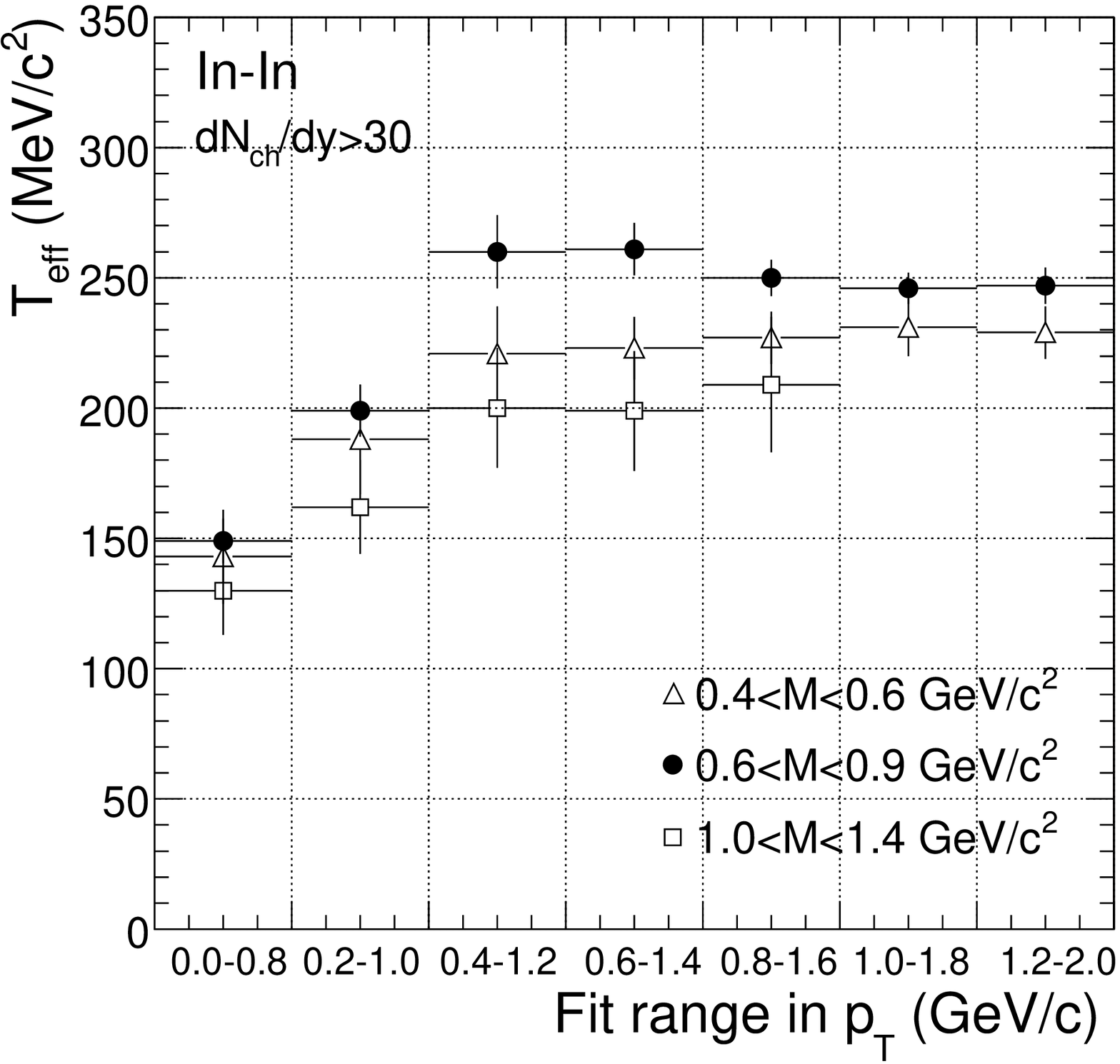}
\caption{Left: Acceptance-corrected $m_{T}$ spectra for three mass
windows, summed over centralities (excluding the peripheral
bin). Right: Inverse slope parameter $\Teff$ from differential fits
of the $m_{T}$ spectra (see text).}
\label{fig5}
\end{figure*}

A differential analysis in $\pT$ was originaly performed and mass spectra associated
with three different $\pT$ windows, without acceptance
correction, were obtained~\cite{hq:2006}. The NA60
acceptance relative to $4\pi$ as a function of $\pT$ as shown in
Fig.~\ref{fig3} (left), implies that one should in principle perform the acceptance correction using a 3-dimensional grid in ($M, \pT, y$) space. This can lead, however, to large errors once the correction is applied. Instead, the correction is performed in 2-dimensional ($M, \pT$) space, using the
measured rapidity distribution as an input. The latter was determined
with an acceptance correction found, in an iterative way, from Monte
Carlo simulations matched to the data in $M$ and $\pT$. On the basis
of this rapidity distribution, 0.1 GeV/$c^2$ bins in M and 0.2
GeV/$c$ bins in $\pT$ were used to determine the remaining
2-dimensional correction. After correction the results were integrated
over the three extended mass windows $0.4<\!\!M\!\!<0.6$,~$0.6<\!\!M\!\!<0.9$
and $1.0\!<\!M\!<\!1.4$ GeV/c$^{2}$. In Fig.~\ref{fig3} (right) the rapidity distribution of the
central mass window is shown for three
different $\pT$ windows,  exhibiting a close resemblance to the distribution of inclusive pion production, as measured by
NA49 for Pb-Pb and NA60 for In-In.
The results for the acceptance-corrected $p_{T}$ spectra are
summarized in Fig.~\ref{fig4}. The errors shown are purely
statistical. Systematic errors arise from the acceptance corrections
including the rapidity distribution used, the subtraction of the
cocktail, and the subtraction of the combinatorial background plus
fake matches. For $p_{T}\!<\!0.5$ GeV/c, the combinatorial background
contributes most, ranging from 10 to 25\% for semiperipheral up to
central. For $p_{T}\!>\!1$ GeV/c, the statistical errors dominate. The
data show a significant dependence on mass, but hardly any on
centrality. To bear out the differences in mass more clearly, the data
were summed over the three more central bins and plotted in Fig.~\ref{fig5} (left)
as a function of $m_T$. The inverse slope parameter $\Teff$ as
determined from differential fits of the $m_{T}$ spectra with
$\exp(-m_{T}/\Teff)$, using a sliding window in $p_{T}$, is plotted
on the right. Instead of flattening at very low $m_{T}$ as should be expected from radial flow, strikingly all spectra \textit{steepen}, which is equivalent to very small
values of $\Teff$. For comparison, the $\phi$ resonance, placed just in between the
upper two mass windows, does flatten as expected. Moreover,
depending on the fit region, $\Teff$ covers an unusually large
dynamic range. Finally, the largest masses have the steepest $m_{T}$
spectrum, \textit{i.e.} the smallest value of $\Teff$ everywhere, again
contrary to radial flow and to what is usually observed for
hadrons. All this suggests that different mass regions are coupled to
basically different emission sources.
%figure 5
%end figure 5
Summarizing, the previously measured excess mass spectra and the new
acceptance-corrected $p_{T}$ and $m_{T}$ spectra present an unexpected behaviour. Beyond the $\rho$
spectral function this may lead to a better understanding
of the continuum part of the spectra for $M\!>1$ GeV/c$^2$, possibly disentangling parton-hadron duality.

\end{document}